\documentclass[preprints,article,accept,oneauthor,pdftex]{mdpi} 
\firstpage{1} 
\pubvolume{1}
\issuenum{1}
\articlenumber{0}
\pubyear{2021}
\copyrightyear{2021}
\datereceived{} 
\dateaccepted{} 
\datepublished{} 
\hreflink{https://doi.org/} 
\pdfoutput=1
\usepackage{braket}
\newcommand{\MyColumnfixparacol}{\end{paracol}}
\Title{Quantum Reservoir Computing for Speckle-Disorder Potentials}
\TitleCitation{Quantum Reservoir Computing for Speckle-Disorder Potentials}

\Author{Pere Mujal $^{1}$\orcidA{}}

\AuthorNames{Pere Mujal}

\AuthorCitation{Mujal, P.}

\address{%
$^{1}$ \quad IFISC, Institut de F\'{\i}sica Interdisciplin\`{a}ria i Sistemes Complexos (UIB-CSIC), UIB Campus, E-07122 Palma, Mallorca, Spain; peremujal@ifisc.uib-csic.es}

\corres{Correspondence: peremujal@ifisc.uib-csic.es}

\abstract{Quantum reservoir computing is a machine-learning approach designed to exploit the dynamics of quantum systems with memory to process information. As an advantage, it presents the possibility to benefit from the quantum resources provided by the reservoir combined with a simple and fast training strategy. In this work, this technique is introduced with a quantum reservoir of spins and it is applied to find the ground-state energy of an additional quantum system. The quantum reservoir computer is trained with a linear model to predict the lowest energy of a particle in the presence of different speckle-disorder potentials. The performance of the task is analyzed with a focus on
the observable quantities extracted from the reservoir and it shows to be enhanced when two-qubit correlations are employed.
}
\begin{document}
\section{Introduction}

In the last few years, the study of quantum systems has taken advantage of the increasing interest and the developments of machine learning techniques to face both theoretical and experimental challenges, which has lead to the emergence of the broad field of quantum machine learning~\cite{Biamonte2017,Dunjko_2018,doi:10.1080/23746149.2020.1797528,doi:10.1080/00107514.2014.964942,RevModPhys.91.045002}. Some successful examples of the use of machine learning include, among others, the detection and classification of quantum phases~\cite{PhysRevLett.125.170603,PhysRevB.99.121104,Carrasquilla2017,PhysRevB.94.195105,Rem2019,PhysRevB.97.134109,PhysRevB.100.045129,PhysRevB.95.245134}, the prediction of the ground-state energy and other characteristic quantities of quantum systems~\cite{PhysRevA.96.042113,PhysRevE.101.063308,Pilati2019,10.21468/SciPostPhys.10.3.073}, and the enhanced control and readout in experimental setups~\cite{Wigley2016,Tranter2018,Barker_2020,PhysRevX.10.011006}. Additionally, many efforts are devoted to develop machine-learning algorithms that exploit quantum resources, aiming to find a quantum advantage in performing tasks. Quantum reservoir computing (QRC) and related approaches belong to this last category~\cite{mujal2021opportunities,NakajimaGoshReview}.

The concept of QRC was introduced in \cite{Nakajima2017} as an extension to the quantum realm of classical reservoir computing (RC)~\cite{lukovsevivcius2009reservoir,jaeger2004harnessing,maass2004computational,brunner2019photonic}. The main idea behind RC, as an unconventional computing method~\cite{konkoli2017reservoir,Adamatzky}, is the use of the natural dynamics of systems to process information, together with a simplified training strategy~\cite{butcher2013reservoir}. For supervised learning techniques, for instance in the case of deep neural networks, one of the major drawbacks is the training process of models with typically thousands of free parameters to be optimally adjusted, which requires a lot of computational resources and/or time. Instead of that, in RC, the connections between the constituents of the reservoir are kept fixed and only the output quantities from the reservoir are involved in the training process and could be easily retrained for a different purpose. This scheme has shown to be sufficient to achieve very good performances in diverse tasks~\cite{lukovsevivcius2012practical,antonik2019human,alfaras2019fast,pathak2018model}.

Quantum reservoirs are good candidates to be exploited for computational purposes for several reasons. First of all, the number of degrees of freedom in quantum systems increases exponentially with the number of constituents. Therefore, with relatively small systems, a large state space is available, which has been shown to be beneficial, i.e. it increases the memory capacity~\cite{schuld2019quantum,Abbas2021,Nakajima2017,nokkala2021high,martinez2020information}. In second place, the presence of entanglement can also contribute to achieve a quantum advantage when quantum correlations are exploited~\cite{martinez2020information}.
Finally, there exist several proposals suitable to be implemented in a wide variety of experimental platforms to realize not only classical tasks but also quantum ones~\cite{mujal2021opportunities}, for instance, entanglement detection~\cite{Ghosh2019}, quantum state
tomography~\cite{9153954} and quantum state preparation~\cite{ghosh2019neuromorphic,KRISNANDA2021141}.

In this article, the possibility of using a quantum reservoir to study another quantum system is explored as an alternative to classical machine-learning models. A first goal in this work is to show that a quantum reservoir can be used to make predictions on the ground-state energy of a quantum particle in a speckle-disorder potential~\cite{aspect2009anderson} by only providing as an input this external potential and by using only a linear model to train the output observables.
This problem is of relevance for understanding the Anderson localization phenomenon in quantum systems due to the presence of disorder, which determines their transport properties~\cite{PhysRev.109.1492}.
Additionally, we aim to analyze the effect on the performance when two-body quantum correlations in the reservoir are used compared with one-body observables for the mentioned task.

This work is organized as follows. In Section \ref{Sec2}, the details on the quantum system in study are provided together with the database used. The description of how the input is encoded into the quantum reservoir is found in Section \ref{Sec3}. In Section \ref{Sec4}, the characteristics of the quantum reservoir system are explained. In Section \ref{Sec5}, the quantum reservoir computing procedure is presented with the mathematical description of the state of the reservoir and its observables. The expressions of the trained models and an analysis of their performance are given in Section \ref{Sec6}. Finally, the discussion of the results and the conclusions are in Section \ref{Sec7}.
\section{Database of Speckle-Disorder Potentials and Ground-State Energies}
\label{Sec2}
The problem to be addressed with the model proposed in this work consists in finding the 
lowest energy of the following Hamiltonian, which describes a particle of mass $m$ in one dimension with position $x$ in the presence of an external potential $V(x)$:

\begin{equation}
\label{Eqhamiltonian}
\mathcal{H}=-\frac{\hbar^2}{2m}\frac{\partial^2}{\partial x^2}+V(x),
\end{equation}

where $\hbar$ is the reduced Planck constant. $V(x)$ is a speckle potential that in cold-atoms experiments is created by means of optical fields passing through a diffusion plate~\cite{Clement_2006,billy2008direct,roati2008anderson} and can be numerically produced with Gaussian random numbers~\cite{PhysRevA.73.013606,ley1989specklhunte}. This potential introduces disorder into the system and originates the Anderson localization phenomenon~\cite{PhysRev.109.1492,aspect2009anderson}.

In previous studies, classical machine-learning models with convolutional neural networks have shown to be able to make very good predictions on the first energies of this system and for different system sizes by applying transfer-learning protocols~\cite{Pilati2019}. Additionally, the extension to the system with few repulsively interacting bosons~\cite{PhysRevA.100.013603} has also been explored including the particle number as an additional feature to the trained model~\cite{10.21468/SciPostPhys.10.3.073}.

The database used in this article is part of the database used in ~\cite{10.21468/SciPostPhys.10.3.073}, which is publicly available at ~\cite{zenodo}. In this work, the first 10000 speckle-potential instances of the single-particle dataset and their corresponding ground-state energies are used. The energies in the database were computed numerically by means of exact diagonalization as explained in \cite{PhysRevA.100.013603} and in more detail in \cite{pmtphdthesis}.
\section{Input Ecoding into the Quantum Reservoir}
\label{Sec3}
The values of each different speckle-potential instance in the dataset are provided in a discrete grid in space of $K=1024$ points, $V(x_k)=V_k$, with $k=1, \, ... \, ,K$. Therefore, our input is a set of vectors of size $1024$ where the spatial structure of each potential is given by the order of the elements. For this reason, the values of the potential are introduced into the dynamics of the quantum reservoir in the same order. From the point of view of the quantum reservoir, a given speckle potential corresponds to an external time-dependent signal fed at discrete times, $t=k\Delta t$.

\begin{figure}[t!]
\includegraphics[width=\columnwidth]{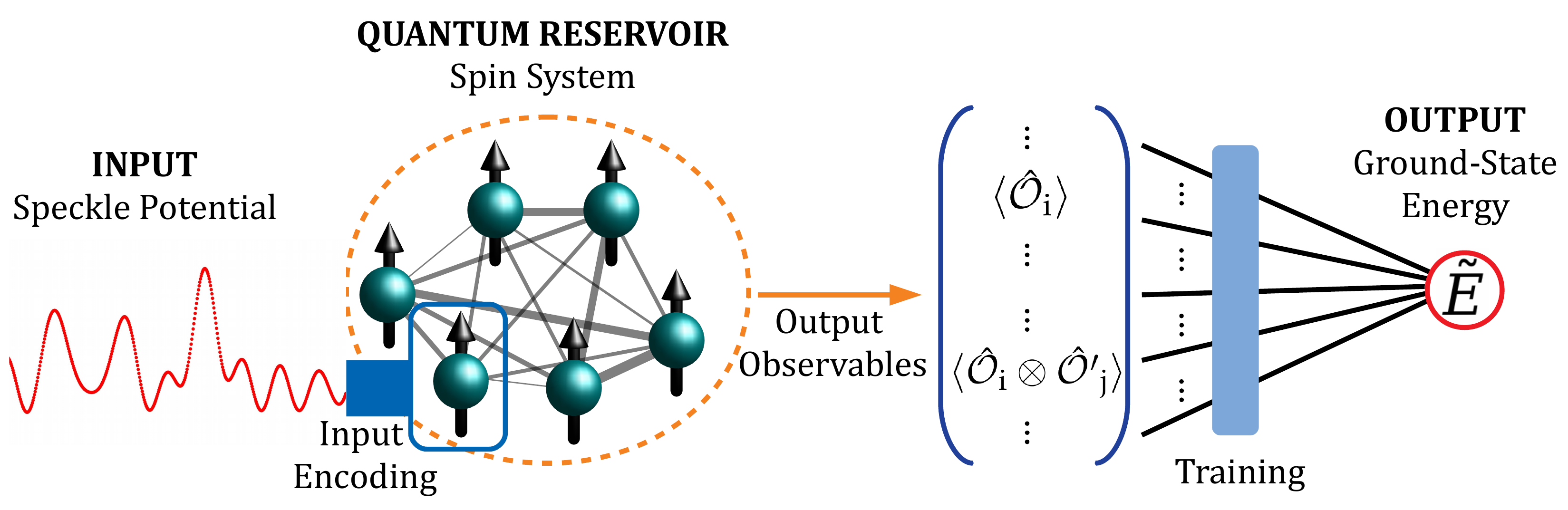}
\caption{Schematic representation of the use of a quantum reservoir to predict the ground-state energy of a quantum particle in a speckle potential. The spatial-dependent values of the speckle potential are transformed into a time-dependent signal fed into the state of one spin of the reservoir at discrete time steps. The quantum reservoir system evolves between input injections. Different observables are extracted from the reservoir system and used in the training process to produce ground-state energy predictions.\label{fig1}}
\end{figure}

The input at a given time, $V_k$, is encoded in one of the qubits of the system (See Figure~\ref{fig1}), qubit 1, by setting its state as~\cite{martinez2021dynamical,Nakajima2017,Kutvonen2020,martinez2020information}:

\begin{equation}
\label{eqencoding}
\ket{\psi_k}=\sqrt{1-s_k}\ket{0}+\sqrt{s_k}\ket{1},
\end{equation}

where $s_k=V_k/V_{\textnormal{max}}$ and the value of $V_{\textnormal{max}}$ is always the same and fixed as the maximum reached among all the $10000$ speckle instances in the dataset. In this way, $s_k\in[0,1]$ to have a properly normalized state for the qubit and all inputs are re-scaled the same quantity. The basis states that are used are the eigenstates of the Pauli matrix $\hat{\sigma}^z$, namely $\hat{\sigma}^z\ket{0}=\ket{0}$ and $\hat{\sigma}^z\ket{1}=-\ket{1}$.

\section{Hamiltonian of the Reservoir of Spins}
\label{Sec4}
The quantum reservoir employed in this work is a system consisting of $N=6$ spins (or qubits). The unitary dynamics of this system are governed by the following transverse-field Ising Hamiltonian:

\begin{equation}
\hat{H}_{\textnormal{R}}=\frac{1}{2}\sum_{i=1}^{N}h\hat{\sigma}_i^z+\sum_{i<j}^{N}J_{ij}\hat{\sigma}_i^x\hat{\sigma}_j^x,
\label{hamtransverseising}
\end{equation}

where $\hat{\sigma}_i^z$ and $\hat{\sigma}_j^x$ are Pauli matrices acting on qubits $i$ and $j$, respectively.
The spin-spin couplings $J_{ij}$, represented as lines of different thickness in Figure~\ref{fig1}, are randomly generated once from a uniform distribution in the interval $[-J_s/2,J_s/2]$ and then kept constant. We work on a system of units with $\hbar=1$ and $J_s=1$. The time intervals $\Delta t$, are expressed in units of $1/J_s$, and $h$, that corresponds to an external magnetic field in the z direction, is fixed at $h=10 J_s$, such that the system is in the appropriate dynamical regime~\cite{martinez2021dynamical}.
This kind of system was in the original proposal of QRC in~\cite{Nakajima2017} and has been extensively studied for information processing purposes in  further several works~\cite{Chen2019,PhysRevApplied.11.034021,martinez2020information,martinez2021dynamical,Kutvonen2020,Mujal_2021nonlinearities,higherorderqrc}. 

\section{Quantum Reservoir Computer Operation}
\label{Sec5}
The role of the quantum reservoir is to provide a map between the input speckle to the output observables. They carry the information about the input that has been processed during the time-evolution of the reservoir system. The memory of the system, for the present purposes, is exploited within each instance. However, the system is reset before the introduction of each speckle-potential at time $t=0$. In this way, there are no dependencies between consecutive instances.
The general scheme of the procedure is depicted in Figure~\ref{fig1}.

The density matrix that describes the quantum state of the reservoir of spins before the injection of each potential reads:

\begin{equation}
    \rho_{0}=\ket{0,\, ...\,,0}\bra{0,\, ...\,,0}.
\end{equation}

Afterwards, for a given speckle-potential instance, the state of the reservoir at each time step $k$, is given by:

\begin{equation}
    \rho_{k}=e^{-i \hat{H}_{\textnormal{R}} \Delta t}\left(\ket{\psi_k}\bra{\psi_k}\otimes \textnormal{Tr}_1({\rho}_{k-1}) \right) e^{i \hat{H}_{\textnormal{R}} \Delta t},
\end{equation}

where $\textnormal{Tr}_1()$ indicates the partial trace with respect to the first qubit. The dependence on the speckle points, $V_k$, is through fixing the state of the first qubit, $\ket{\psi_k}\bra{\psi_k}$, in the form of Equation~\eqref{eqencoding}. Between input injections, there is a unitary evolution of the state of the reservoir of $\Delta t$ duration governed by the Hamiltonian in \eqref{hamtransverseising}.

From the state of the reservoir, the expectation values of the following single-qubit observables at each time step $k$ are computed as:

\begin{equation}
    \langle \hat{\sigma}^{\alpha}_i \rangle_k = \textnormal{Tr}\left(\rho_k \hat{\sigma}^{\alpha}_i\right),
\end{equation}

for all spins, $i=1, \,....\, N$, and in the three directions $\alpha=x,y,z$. Afterwards, the average over all time steps is taken to obtain:

\begin{equation}
\label{eqsinglemeans}
     \langle \hat{\sigma}^{\alpha}_i \rangle \equiv \frac{1}{K}\sum_{k=1}^{K} \langle \hat{\sigma}^{\alpha}_i \rangle_k.
\end{equation}

Similarly, the expectation values of the two-qubit observables are calculated,

\begin{equation}
    \langle \hat{\sigma}^{\alpha}_i\hat{\sigma}^{\beta}_j \rangle_k = \textnormal{Tr}\left(\rho_k  \hat{\sigma}^{\alpha}_i\hat{\sigma}^{\beta}_j\right),
\end{equation}

and

\begin{equation}
\label{eqtwomeans}
    \langle \hat{\sigma}^{\alpha}_i\hat{\sigma}^{\beta}_j \rangle \equiv \frac{1}{K}\sum_{k=1}^{K} \langle \hat{\sigma}^{\alpha}_i\hat{\sigma}^{\beta}_j \rangle_k,
\end{equation}

with $i<j$ and $\alpha,\beta=x,y,z$.
\section{Training and Predictions of the Models}
\label{Sec6}
From the output observables of the quantum reservoir, two models to make predictions, $\tilde{E}$, on the ground-state energies are proposed. Both are constructed by fitting a  least-squares linear model with the training dataset, which corresponds to the first 7500 speckle instances and target energies $E$. The quality of the models is tested with the remaining 2500 potentials and it is quantified by the mean absolut error (MAE),

\begin{equation}
    \textnormal{MAE}=\frac{1}{M}\sum_{l=1}^M|E_l-\tilde{E}_l|,
\end{equation}

and the coefficient of determination

\begin{equation}
    R^2=1-\frac{\sum_{l=1}^M(E_l-\tilde{E}_l)^2}{\sum_{l=1}^M (\bar{E}-E_l)^2},
\end{equation}

where $\bar{E}$ is the mean energy $\bar{E}\equiv (1/M)\sum_{l=1}^M E_l$, and $M$ is the number of instances, $M=7500$ for the training data, and $M=2500$ for the test data.
If $R^2=0$, the predicted and the target energies are not linearly related, whereas for $R^2=1$ the predictions are perfect.

In the first case, the single-qubit observables in Equation \eqref{eqsinglemeans} are employed
and the final output, $\tilde{E}$, for each instance is written as:

\begin{equation}
\label{modelmeans}
\begin{gathered}
    \tilde{E}=v_0+\sum_{i=1}^{N}\left(
    v_{1,i}\langle \hat{\sigma}^{x}_i \rangle
    +v_{2,i}\langle \hat{\sigma}^{y}_i \rangle
    +v_{3,i}\langle \hat{\sigma}^{z}_i \rangle
    +v_{4,i}\langle \hat{\sigma}^{x}_i \rangle^2
    +v_{5,i}\langle \hat{\sigma}^{y}_i \rangle^2
    +v_{6,i}\langle \hat{\sigma}^{z}_i \rangle^2\right.
\\
   \left.
    +v_{7,i}\langle \hat{\sigma}^{x}_i \rangle \langle \hat{\sigma}^{y}_i \rangle
    +v_{8,i}\langle \hat{\sigma}^{y}_i \rangle \langle \hat{\sigma}^{z}_i \rangle
    +v_{9,i}\langle \hat{\sigma}^{z}_i \rangle \langle \hat{\sigma}^{x}_i \rangle
    +v_{10,i}\langle \hat{\sigma}^{x}_i \rangle \langle \hat{\sigma}^{y}_i \rangle \langle \hat{\sigma}^{z}_i \rangle
    \right).
\end{gathered}
\end{equation}

In this last model and for a system with $N=6$, there are 61 free parameters, $v$ with the corresponding labels, that are optimized during the training.

In the second case, there are 451 different weights $w$ to be adjusted when the two-qubit quantities in Equation \eqref{eqtwomeans} are used in the similar following way: 

\begin{equation}
\label{modelcorrrelations}
\begin{gathered}
    \tilde{E}=w_0+\sum_{i<j=1}^{N}\left(
    w_{1,i,j}\langle \hat{\sigma}^{x}_i  \hat{\sigma}^{x}_j \rangle
    +w_{2,i,j}\langle \hat{\sigma}^{y}_i\hat{\sigma}^{y}_j \rangle
    +w_{3,i,j}\langle \hat{\sigma}^{z}_i\hat{\sigma}^{z}_j \rangle
    \right.
\\
   \left.
    +w_{4,i,j}\langle \hat{\sigma}^{x}_i\hat{\sigma}^{x}_j \rangle^2
    +w_{5,i,j}\langle \hat{\sigma}^{y}_i\hat{\sigma}^{y}_j \rangle^2
    +w_{6,i,j}\langle \hat{\sigma}^{z}_i\hat{\sigma}^{z}_j \rangle^2
    +w_{7,i,j}\langle \hat{\sigma}^{x}_i  \hat{\sigma}^{x}_j \rangle \langle \hat{\sigma}^{y}_i\hat{\sigma}^{y}_j \rangle\right.
\\
   \left.
    +w_{8,i,j}\langle \hat{\sigma}^{y}_i\hat{\sigma}^{y}_j \rangle \langle \hat{\sigma}^{z}_i\hat{\sigma}^{z}_j \rangle
    +w_{9,i,j}\langle \hat{\sigma}^{z}_i\hat{\sigma}^{z}_j \rangle \langle \hat{\sigma}^{x}_i  \hat{\sigma}^{x}_j \rangle
    +w_{10,i,j}\langle \hat{\sigma}^{x}_i  \hat{\sigma}^{x}_j \rangle \langle \hat{\sigma}^{y}_i\hat{\sigma}^{y}_j \rangle \langle \hat{\sigma}^{z}_i\hat{\sigma}^{z}_j \rangle
    \right)
\\
    +\sum_{i\neq j=1}^{N}\left(
    w_{11,i,j}\langle \hat{\sigma}^{x}_i  \hat{\sigma}^{y}_j \rangle
    +w_{12,i,j}\langle \hat{\sigma}^{y}_i\hat{\sigma}^{z}_j \rangle
    +w_{13,i,j}\langle \hat{\sigma}^{z}_i\hat{\sigma}^{x}_j \rangle
    \right.
\\
   \left.
    +w_{14,i,j}\langle \hat{\sigma}^{x}_i\hat{\sigma}^{y}_j \rangle^2
    +w_{15,i,j}\langle \hat{\sigma}^{y}_i\hat{\sigma}^{z}_j \rangle^2
    +w_{16,i,j}\langle \hat{\sigma}^{z}_i\hat{\sigma}^{x}_j \rangle^2
    +w_{17,i,j}\langle \hat{\sigma}^{x}_i  \hat{\sigma}^{y}_j \rangle \langle \hat{\sigma}^{y}_i\hat{\sigma}^{z}_j \rangle\right.
\\
   \left.
    +w_{18,i,j}\langle \hat{\sigma}^{y}_i\hat{\sigma}^{z}_j \rangle \langle \hat{\sigma}^{z}_i\hat{\sigma}^{x}_j \rangle
    +w_{19,i,j}\langle \hat{\sigma}^{z}_i\hat{\sigma}^{x}_j \rangle \langle \hat{\sigma}^{x}_i  \hat{\sigma}^{y}_j \rangle
    +w_{20,i,j}\langle \hat{\sigma}^{x}_i  \hat{\sigma}^{y}_j \rangle \langle \hat{\sigma}^{y}_i\hat{\sigma}^{z}_j \rangle \langle \hat{\sigma}^{z}_i\hat{\sigma}^{x}_j \rangle
    \right).
\end{gathered}
\end{equation}

\begin{figure}[t!]
\includegraphics[width=0.48\columnwidth]{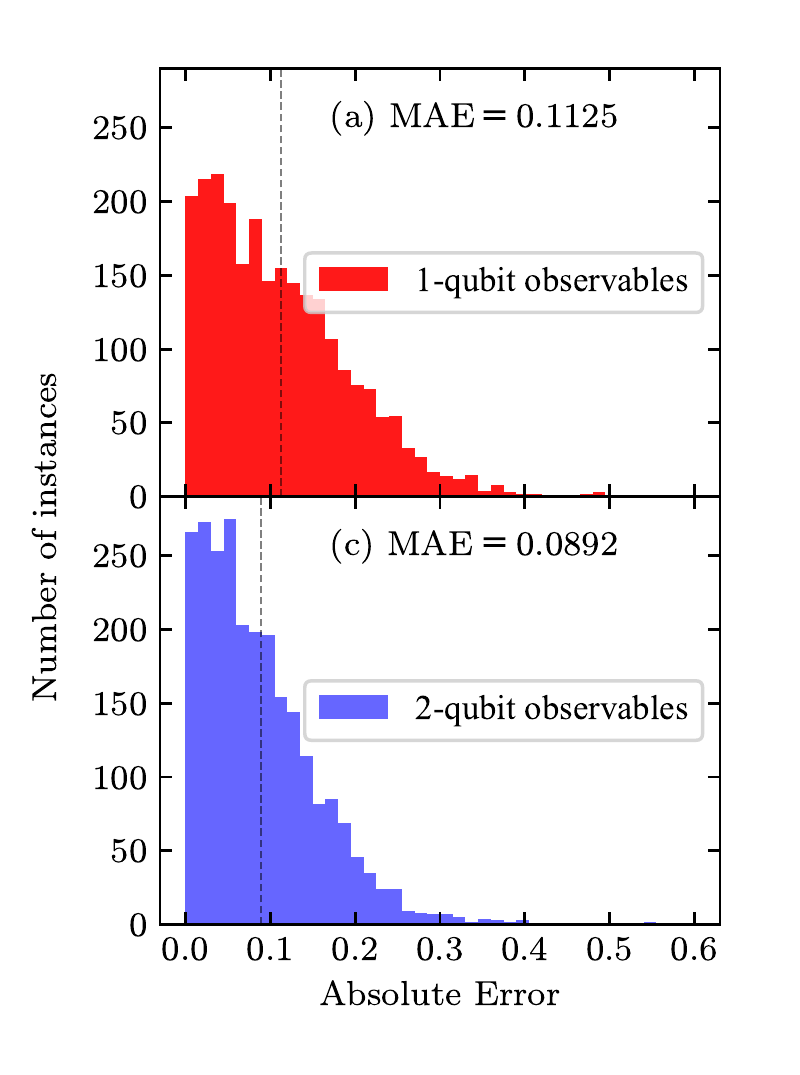}
\includegraphics[width=0.48\columnwidth]{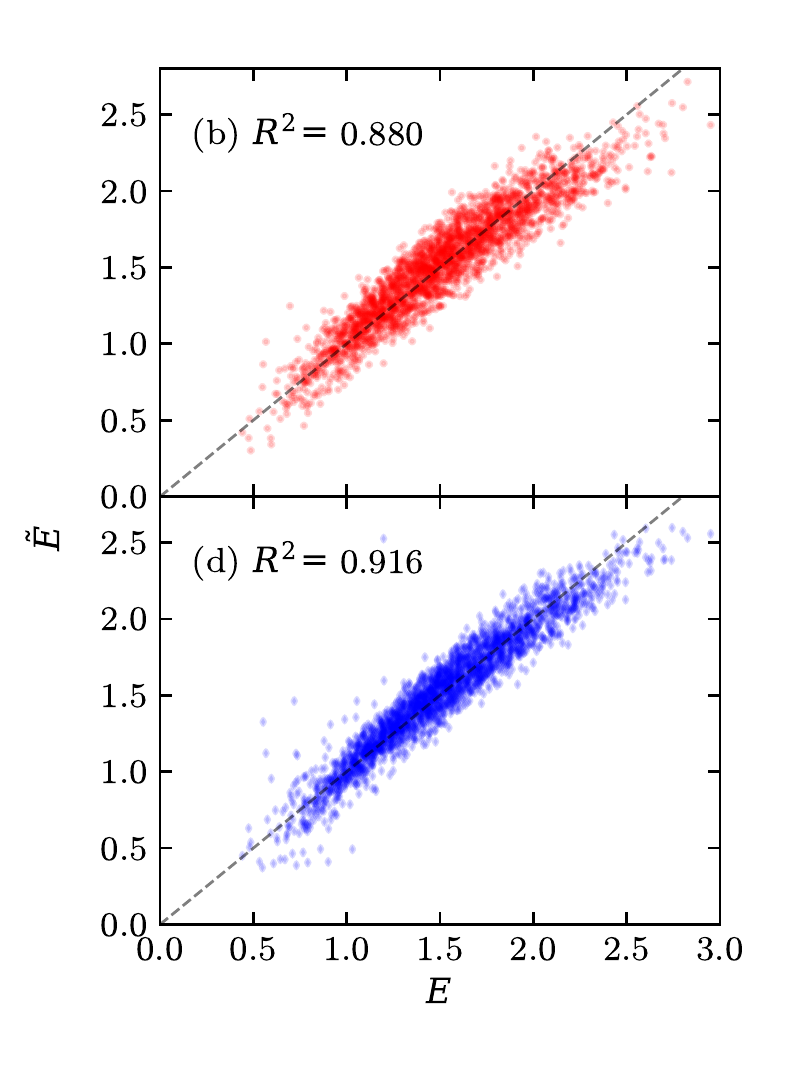}
\caption{(Left) Absolut error distributions corresponding to the test data of the model with single-qubit observables in (a) and two-qubit observables in (c). A dashed line indicates the mean absolut error (MAE). (Right) Predicted energies $\tilde{E}$ as a function of the target energies $E$ for the test data. In (b) the predictions are made with the model in Equation~\eqref{modelmeans} and in (d) with the one in ~\eqref{modelcorrrelations}. The ideal situation, with perfect predictions, is depicted with a dashed line and would correspond to $R^2=1$.\label{figperformance}}
\end{figure}
As in the neural-network models in~\cite{Pilati2019} and~\cite{10.21468/SciPostPhys.10.3.073}, the predictions $\tilde{E}$ are functions of the speckle points in space. In the present case, the required
nonlinear dependence of the outputs on the input values of the potential are, in general, guaranteed by both the form of input encoding in Equation~\eqref{eqencoding} and an appropriate choice of the Hamiltonian parameters~\cite{Mujal_2021nonlinearities}. Beyond that, further nonlinear dependencies on the input have been introduced by combining the observables in \eqref{modelmeans} and \eqref{modelcorrrelations} because it is beneficial to increase the performance without losing the linearity of the models.

The quality of the predictions of the two models is shown in Figure~\ref{figperformance}. Remarkably, the model that produces the predictions from the single-qubit observables of the reservoir with only 61 optimized parameters is able to learn from the training data without overfitting. The MAE and $R^2$ for the training data are 0.1126 and 0.8808, respectively, and practically equal to the values of the test data provided in Figure~\ref{figperformance}. In panel (a), in the distribution of the absolute error, there is a considerable number of speckle instances whose error is below the MAE. Moreover in panel (b), there is a clear correlation between the target energies and the predicted ones for the test data reflected on the large value of $R^2=0.880$, which is close to 0.9. If the model of two-qubit observables is employed instead, the accuracy of the predictions is, in general, improved. The peak of the distribution in panel (c) of the absolute error is sharper and closer to 0, as well as the value of the MAE. In accordance, in the scatter plot in panel (d) the value of $R^2$ surpasses 0.9 and we are closer to the ideal situation. Also in this case with much more free parameters, 451, the comparison between the values of the MAE$=0.0833$ and $R^2=0.936$ for the training data and the test data in panels (c) and (d) indicates that there is not a significant overfitting problem.
%
\section{Discussion and Conclusions}
\label{Sec7}
The results obtained with the models proposed in this work show that a quantum reservoir is suitable to be used to address the problem of making predictions on the ground-state energy of different speckle-disorder potentials.
By following this approach, the computational capabilities of the quantum dynamics of the system are exploited and linear models with the observables of the quantum reservoir are sufficient to achieve a noticeable accuracy. In this way, we have taken advantage of both a simple and fast training strategy and the presence of quantum correlations.
This paves the way to further develop models that follow a similar strategy, for instance, for systems of interacting particles in the presence of a speckle potential.

In fact, for practical purposes, the quality of the predictions should be increased in order to compete with state-of-the-art deep convolutional neural networks. To reach this aim, several strategies could be followed.
It would be interesting to explore the effect of changing the form of the input encoding into the quantum reservoir and to study its impact on the quality of the predictions of the models. In addition to that, in the present work we have only explored the possibility of using single-qubit and two-qubit observables. The extension to three-body quantities and beyond should be considered and could contribute to improve the performance. This would lead to more flexible models, as the number of observables would be increased as well as the number of free parameters. Additionally, in a similar way, increasing the number of spins would enlarge the Hilbert space and increase the capabilities of the quantum reservoir. Regarding the Hamiltonian of the reservoir system, the values of the couplings and the external magnetic field could be seen as hyperparameters. As the performance in realizing the task depends on their values, to improve the results presented in this work, they could be optimized by defining an additional validation dataset.
\vspace{6pt}

\funding{This research was funded by: the Spanish State Research Agency, through the Severo Ochoa and Mar\'ia de Maeztu Program for Centers and Units of Excellence in R\&D, grant number MDM-2017-0711; Comunitat Autònoma de les Illes Balears, Govern de les Illes Balears, through the QUAREC project, grant number PRD2018/47; the Spanish State Research Agency through the  QUARESC project, grant numbers PID2019-109094GB-C21 and -C22/AEI/10.13039/501100011033.}

\dataavailability{The data used in this study are openly available in Zenodo at \cite{zenodo}.}

\acknowledgments{The author acknowledges Rodrigo Martínez-Peña, Gian Luca Giorgi, Miguel C. Soriano and Roberta Zambrini for useful discussions and valuable comments and a careful reading of this manuscript.}
\conflictsofinterest{The author declares no conflict of interest. The funders had no role in the design of the study; in the collection, analyses, or interpretation of data; in the writing of the manuscript, or in the decision to publish the results.}
%
\abbreviations{Abbreviations}{
The following abbreviations are used in this manuscript:\\

\noindent 
\begin{tabular}{@{}ll}
QRC & Quantum Reservoir Computing\\
RC & Reservoir Computing\\
MAE & Mean Absolute Error
\end{tabular}}

\MyColumnfixparacol
%
\reftitle{References}

%

%

\begin{thebibliography}{999}

\bibitem[Biamonte \em{et~al.}(2017)Biamonte, Wittek, Pancotti, Rebentrost,
  Wiebe, and Lloyd]{Biamonte2017}
Biamonte, J.; Wittek, P.; Pancotti, N.; Rebentrost, P.; Wiebe, N.; Lloyd, S.
\newblock Quantum machine learning.
\newblock {\em Nature} {\bf 2017}, {\em 549},~195--202.
\newblock
  doi:{\changeurlcolor{black}\href{https://doi.org/10.1038/nature23474}{\detokenize{10.1038/nature23474}}}.

\bibitem[Dunjko and Briegel(2018)]{Dunjko_2018}
Dunjko, V.; Briegel, H.J.
\newblock Machine learning {\&} artificial intelligence in the quantum domain:
  a review of recent progress.
\newblock {\em Reports on Progress in Physics} {\bf 2018}, {\em 81},~074001.
\newblock
  doi:{\changeurlcolor{black}\href{https://doi.org/10.1088/1361-6633/aab406}{\detokenize{10.1088/1361-6633/aab406}}}.

\bibitem[Carrasquilla(2020)]{doi:10.1080/23746149.2020.1797528}
Carrasquilla, J.
\newblock Machine learning for quantum matter.
\newblock {\em Advances in Physics: X} {\bf 2020}, {\em 5},~1797528.
\newblock
  doi:{\changeurlcolor{black}\href{https://doi.org/10.1080/23746149.2020.1797528}{\detokenize{10.1080/23746149.2020.1797528}}}.

\bibitem[Schuld \em{et~al.}(2015)Schuld, Sinayskiy, and
  Petruccione]{doi:10.1080/00107514.2014.964942}
Schuld, M.; Sinayskiy, I.; Petruccione, F.
\newblock An introduction to quantum machine learning.
\newblock {\em Contemporary Physics} {\bf 2015}, {\em 56},~172--185.
\newblock
  doi:{\changeurlcolor{black}\href{https://doi.org/10.1080/00107514.2014.964942}{\detokenize{10.1080/00107514.2014.964942}}}.

\bibitem[Carleo \em{et~al.}(2019)Carleo, Cirac, Cranmer, Daudet, Schuld,
  Tishby, Vogt-Maranto, and Zdeborov\'a]{RevModPhys.91.045002}
Carleo, G.; Cirac, I.; Cranmer, K.; Daudet, L.; Schuld, M.; Tishby, N.;
  Vogt-Maranto, L.; Zdeborov\'a, L.
\newblock Machine learning and the physical sciences.
\newblock {\em Rev. Mod. Phys.} {\bf 2019}, {\em 91},~045002.
\newblock
  doi:{\changeurlcolor{black}\href{https://doi.org/10.1103/RevModPhys.91.045002}{\detokenize{10.1103/RevModPhys.91.045002}}}.

\bibitem[Kottmann \em{et~al.}(2020)Kottmann, Huembeli, Lewenstein, and
  Ac\'{\i}n]{PhysRevLett.125.170603}
Kottmann, K.; Huembeli, P.; Lewenstein, M.; Ac\'{\i}n, A.
\newblock Unsupervised Phase Discovery with Deep Anomaly Detection.
\newblock {\em Phys. Rev. Lett.} {\bf 2020}, {\em 125},~170603.
\newblock
  doi:{\changeurlcolor{black}\href{https://doi.org/10.1103/PhysRevLett.125.170603}{\detokenize{10.1103/PhysRevLett.125.170603}}}.

\bibitem[Dong \em{et~al.}(2019)Dong, Pollmann, and Zhang]{PhysRevB.99.121104}
Dong, X.Y.; Pollmann, F.; Zhang, X.F.
\newblock Machine learning of quantum phase transitions.
\newblock {\em Phys. Rev. B} {\bf 2019}, {\em 99},~121104.
\newblock
  doi:{\changeurlcolor{black}\href{https://doi.org/10.1103/PhysRevB.99.121104}{\detokenize{10.1103/PhysRevB.99.121104}}}.

\bibitem[Carrasquilla and Melko(2017)]{Carrasquilla2017}
Carrasquilla, J.; Melko, R.G.
\newblock Machine learning phases of matter.
\newblock {\em Nature Physics} {\bf 2017}, {\em 13},~431--434.
\newblock
  doi:{\changeurlcolor{black}\href{https://doi.org/10.1038/nphys4035}{\detokenize{10.1038/nphys4035}}}.

\bibitem[Wang(2016)]{PhysRevB.94.195105}
Wang, L.
\newblock Discovering phase transitions with unsupervised learning.
\newblock {\em Phys. Rev. B} {\bf 2016}, {\em 94},~195105.
\newblock
  doi:{\changeurlcolor{black}\href{https://doi.org/10.1103/PhysRevB.94.195105}{\detokenize{10.1103/PhysRevB.94.195105}}}.

\bibitem[Rem \em{et~al.}(2019)Rem, K{\"a}ming, Tarnowski, Asteria,
  Fl{\"a}schner, Becker, Sengstock, and Weitenberg]{Rem2019}
Rem, B.S.; K{\"a}ming, N.; Tarnowski, M.; Asteria, L.; Fl{\"a}schner, N.;
  Becker, C.; Sengstock, K.; Weitenberg, C.
\newblock Identifying quantum phase transitions using artificial neural
  networks on experimental data.
\newblock {\em Nature Physics} {\bf 2019}, {\em 15},~917--920.
\newblock
  doi:{\changeurlcolor{black}\href{https://doi.org/10.1038/s41567-019-0554-0}{\detokenize{10.1038/s41567-019-0554-0}}}.

\bibitem[Huembeli \em{et~al.}(2018)Huembeli, Dauphin, and
  Wittek]{PhysRevB.97.134109}
Huembeli, P.; Dauphin, A.; Wittek, P.
\newblock Identifying quantum phase transitions with adversarial neural
  networks.
\newblock {\em Phys. Rev. B} {\bf 2018}, {\em 97},~134109.
\newblock
  doi:{\changeurlcolor{black}\href{https://doi.org/10.1103/PhysRevB.97.134109}{\detokenize{10.1103/PhysRevB.97.134109}}}.

\bibitem[Canabarro \em{et~al.}(2019)Canabarro, Fanchini, Malvezzi, Pereira, and
  Chaves]{PhysRevB.100.045129}
Canabarro, A.; Fanchini, F.F.; Malvezzi, A.L.; Pereira, R.; Chaves, R.
\newblock Unveiling phase transitions with machine learning.
\newblock {\em Phys. Rev. B} {\bf 2019}, {\em 100},~045129.
\newblock
  doi:{\changeurlcolor{black}\href{https://doi.org/10.1103/PhysRevB.100.045129}{\detokenize{10.1103/PhysRevB.100.045129}}}.

\bibitem[Schindler \em{et~al.}(2017)Schindler, Regnault, and
  Neupert]{PhysRevB.95.245134}
Schindler, F.; Regnault, N.; Neupert, T.
\newblock Probing many-body localization with neural networks.
\newblock {\em Phys. Rev. B} {\bf 2017}, {\em 95},~245134.
\newblock
  doi:{\changeurlcolor{black}\href{https://doi.org/10.1103/PhysRevB.95.245134}{\detokenize{10.1103/PhysRevB.95.245134}}}.

\bibitem[Mills \em{et~al.}(2017)Mills, Spanner, and
  Tamblyn]{PhysRevA.96.042113}
Mills, K.; Spanner, M.; Tamblyn, I.
\newblock Deep learning and the Schr\"odinger equation.
\newblock {\em Phys. Rev. A} {\bf 2017}, {\em 96},~042113.
\newblock
  doi:{\changeurlcolor{black}\href{https://doi.org/10.1103/PhysRevA.96.042113}{\detokenize{10.1103/PhysRevA.96.042113}}}.

\bibitem[Pilati and Pieri(2020)]{PhysRevE.101.063308}
Pilati, S.; Pieri, P.
\newblock Simulating disordered quantum Ising chains via dense and sparse
  restricted Boltzmann machines.
\newblock {\em Phys. Rev. E} {\bf 2020}, {\em 101},~063308.
\newblock
  doi:{\changeurlcolor{black}\href{https://doi.org/10.1103/PhysRevE.101.063308}{\detokenize{10.1103/PhysRevE.101.063308}}}.

\bibitem[Pilati and Pieri(2019)]{Pilati2019}
Pilati, S.; Pieri, P.
\newblock Supervised machine learning of ultracold atoms with speckle disorder.
\newblock {\em Scientific Reports} {\bf 2019}, {\em 9},~5613.
\newblock
  doi:{\changeurlcolor{black}\href{https://doi.org/10.1038/s41598-019-42125-w}{\detokenize{10.1038/s41598-019-42125-w}}}.

\bibitem[Mujal \em{et~al.}(2021)Mujal, Àlex Martínez~Miguel, Polls,
  Juliá-Díaz, and Pilati]{10.21468/SciPostPhys.10.3.073}
Mujal, P.; Àlex Martínez~Miguel.; Polls, A.; Juliá-Díaz, B.; Pilati, S.
\newblock {Supervised learning of few dirty bosons with variable particle
  number}.
\newblock {\em SciPost Phys.} {\bf 2021}, {\em 10},~73.
\newblock
  doi:{\changeurlcolor{black}\href{https://doi.org/10.21468/SciPostPhys.10.3.073}{\detokenize{10.21468/SciPostPhys.10.3.073}}}.

\bibitem[Wigley \em{et~al.}(2016)Wigley, Everitt, van~den Hengel, Bastian,
  Sooriyabandara, McDonald, Hardman, Quinlivan, Manju, Kuhn, Petersen, Luiten,
  Hope, Robins, and Hush]{Wigley2016}
Wigley, P.B.; Everitt, P.J.; van~den Hengel, A.; Bastian, J.W.; Sooriyabandara,
  M.A.; McDonald, G.D.; Hardman, K.S.; Quinlivan, C.D.; Manju, P.; Kuhn,
  C.C.N.; Petersen, I.R.; Luiten, A.N.; Hope, J.J.; Robins, N.P.; Hush, M.R.
\newblock Fast machine-learning online optimization of ultra-cold-atom
  experiments.
\newblock {\em Scientific Reports} {\bf 2016}, {\em 6},~25890.
\newblock
  doi:{\changeurlcolor{black}\href{https://doi.org/10.1038/srep25890}{\detokenize{10.1038/srep25890}}}.

\bibitem[Tranter \em{et~al.}(2018)Tranter, Slatyer, Hush, Leung, Everett, Paul,
  Vernaz-Gris, Lam, Buchler, and Campbell]{Tranter2018}
Tranter, A.D.; Slatyer, H.J.; Hush, M.R.; Leung, A.C.; Everett, J.L.; Paul,
  K.V.; Vernaz-Gris, P.; Lam, P.K.; Buchler, B.C.; Campbell, G.T.
\newblock Multiparameter optimisation of a magneto-optical trap using deep
  learning.
\newblock {\em Nature Communications} {\bf 2018}, {\em 9},~4360.
\newblock
  doi:{\changeurlcolor{black}\href{https://doi.org/10.1038/s41467-018-06847-1}{\detokenize{10.1038/s41467-018-06847-1}}}.

\bibitem[Barker \em{et~al.}(2020)Barker, Style, Luksch, Sunami, Garrick, Hill,
  Foot, and Bentine]{Barker_2020}
Barker, A.J.; Style, H.; Luksch, K.; Sunami, S.; Garrick, D.; Hill, F.; Foot,
  C.J.; Bentine, E.
\newblock Applying machine learning optimization methods to the production of a
  quantum gas.
\newblock {\em Machine Learning: Science and Technology} {\bf 2020}, {\em
  1},~015007.
\newblock
  doi:{\changeurlcolor{black}\href{https://doi.org/10.1088/2632-2153/ab6432}{\detokenize{10.1088/2632-2153/ab6432}}}.

\bibitem[Flurin \em{et~al.}(2020)Flurin, Martin, Hacohen-Gourgy, and
  Siddiqi]{PhysRevX.10.011006}
Flurin, E.; Martin, L.S.; Hacohen-Gourgy, S.; Siddiqi, I.
\newblock Using a Recurrent Neural Network to Reconstruct Quantum Dynamics of a
  Superconducting Qubit from Physical Observations.
\newblock {\em Phys. Rev. X} {\bf 2020}, {\em 10},~011006.
\newblock
  doi:{\changeurlcolor{black}\href{https://doi.org/10.1103/PhysRevX.10.011006}{\detokenize{10.1103/PhysRevX.10.011006}}}.

\bibitem[Mujal \em{et~al.}(2021)Mujal, Mart{\'\i}nez-Pe{\~n}a, Nokkala,
  Garc{\'\i}a-Beni, Giorgi, Soriano, and Zambrini]{mujal2021opportunities}
Mujal, P.; Mart{\'\i}nez-Pe{\~n}a, R.; Nokkala, J.; Garc{\'\i}a-Beni, J.;
  Giorgi, G.L.; Soriano, M.C.; Zambrini, R.
\newblock Opportunities in Quantum Reservoir Computing and Extreme Learning
  Machines.
\newblock {\em Advanced Quantum Technologies} {\bf 2021}, p. 2100027.
\newblock
  doi:{\changeurlcolor{black}\href{https://doi.org/https://doi.org/10.1002/qute.202100027}{\detokenize{https://doi.org/10.1002/qute.202100027}}}.

\bibitem[Ghosh \em{et~al.}(2021)Ghosh, Nakajima, Krisnanda, Fujii, and
  Liew]{NakajimaGoshReview}
Ghosh, S.; Nakajima, K.; Krisnanda, T.; Fujii, K.; Liew, T.C.H.
\newblock Quantum Neuromorphic Computing with Reservoir Computing Networks.
\newblock {\em Advanced Quantum Technologies} {\bf 2021}, {\em 4},~2100053.
\newblock
  doi:{\changeurlcolor{black}\href{https://doi.org/https://doi.org/10.1002/qute.202100053}{\detokenize{https://doi.org/10.1002/qute.202100053}}}.

\bibitem[Fujii and Nakajima(2017)]{Nakajima2017}
Fujii, K.; Nakajima, K.
\newblock Harnessing Disordered-Ensemble Quantum Dynamics for Machine Learning.
\newblock {\em Phys. Rev. Applied} {\bf 2017}, {\em 8},~024030.
\newblock
  doi:{\changeurlcolor{black}\href{https://doi.org/10.1103/PhysRevApplied.8.024030}{\detokenize{10.1103/PhysRevApplied.8.024030}}}.

\bibitem[Luko{\v{s}}evi{\v{c}}ius and
  Jaeger(2009)]{lukovsevivcius2009reservoir}
Luko{\v{s}}evi{\v{c}}ius, M.; Jaeger, H.
\newblock Reservoir computing approaches to recurrent neural network training.
\newblock {\em Comput. Sci. Rev.} {\bf 2009}, {\em 3},~127--149.
\newblock
  doi:{\changeurlcolor{black}\href{https://doi.org/https://doi.org/10.1016/j.cosrev.2009.03.005}{\detokenize{https://doi.org/10.1016/j.cosrev.2009.03.005}}}.

\bibitem[Jaeger and Haas(2004)]{jaeger2004harnessing}
Jaeger, H.; Haas, H.
\newblock Harnessing Nonlinearity: Predicting Chaotic Systems and Saving Energy
  in Wireless Communication.
\newblock {\em Science} {\bf 2004}, {\em 304},~78--80.
\newblock
  doi:{\changeurlcolor{black}\href{https://doi.org/10.1126/science.1091277}{\detokenize{10.1126/science.1091277}}}.

\bibitem[Maass and Markram(2004)]{maass2004computational}
Maass, W.; Markram, H.
\newblock On the computational power of circuits of spiking neurons.
\newblock {\em Journal of Computer and System Sciences} {\bf 2004}, {\em
  69},~593--616.
\newblock
  doi:{\changeurlcolor{black}\href{https://doi.org/https://doi.org/10.1016/j.jcss.2004.04.001}{\detokenize{https://doi.org/10.1016/j.jcss.2004.04.001}}}.

\bibitem[Brunner \em{et~al.}(2019)Brunner, Soriano, and Van~der
  Sande]{brunner2019photonic}
Brunner, D.; Soriano, M.C.; Van~der Sande, G.
\newblock {\em Photonic Reservoir Computing: Optical Recurrent Neural
  Networks}; Walter de Gruyter GmbH \& Co KG,  2019.

\bibitem[Konkoli(2017)]{konkoli2017reservoir}
Konkoli, Z.
\newblock On reservoir computing: from mathematical foundations to
  unconventional applications. In {\em Advances in unconventional computing};
  Springer,  2017; pp. 573--607.

\bibitem[Adamatzky \em{et~al.}(2007)Adamatzky, Bull, De~Lacy~Costello, Stepney,
  and Teuscher]{Adamatzky}
Adamatzky, A.; Bull, L.; De~Lacy~Costello, B.; Stepney, S.; Teuscher, C.
\newblock {\em Unconventional Computing 2007}; Luniver Press,  2007.

\bibitem[Butcher \em{et~al.}(2013)Butcher, Verstraeten, Schrauwen, Day, and
  Haycock]{butcher2013reservoir}
Butcher, J.B.; Verstraeten, D.; Schrauwen, B.; Day, C.R.; Haycock, P.W.
\newblock Reservoir computing and extreme learning machines for non-linear
  time-series data analysis.
\newblock {\em Neural Networks} {\bf 2013}, {\em 38},~76--89.
\newblock
  doi:{\changeurlcolor{black}\href{https://doi.org/https://doi.org/10.1016/j.neunet.2012.11.011}{\detokenize{https://doi.org/10.1016/j.neunet.2012.11.011}}}.

\bibitem[Luko{\v{s}}evi{\v{c}}ius(2012)]{lukovsevivcius2012practical}
Luko{\v{s}}evi{\v{c}}ius, M.
\newblock A practical guide to applying echo state networks. In {\em Neural
  networks: Tricks of the trade}; Springer,  2012; pp. 659--686.

\bibitem[Antonik \em{et~al.}(2019)Antonik, Marsal, Brunner, and
  Rontani]{antonik2019human}
Antonik, P.; Marsal, N.; Brunner, D.; Rontani, D.
\newblock Human action recognition with a large-scale brain-inspired photonic
  computer.
\newblock {\em Nature Machine Intelligence} {\bf 2019}, {\em 1},~530--537.
\newblock
  doi:{\changeurlcolor{black}\href{https://doi.org/10.1038/s42256-019-0110-8}{\detokenize{10.1038/s42256-019-0110-8}}}.

\bibitem[Alfaras \em{et~al.}(2019)Alfaras, Soriano, and
  Ort{\'\i}n]{alfaras2019fast}
Alfaras, M.; Soriano, M.C.; Ort{\'\i}n, S.
\newblock A Fast Machine Learning Model for ECG-Based Heartbeat Classification
  and Arrhythmia Detection.
\newblock {\em Frontiers in Physics} {\bf 2019}, {\em 7},~103.
\newblock
  doi:{\changeurlcolor{black}\href{https://doi.org/10.3389/fphy.2019.00103}{\detokenize{10.3389/fphy.2019.00103}}}.

\bibitem[Pathak \em{et~al.}(2018)Pathak, Hunt, Girvan, Lu, and
  Ott]{pathak2018model}
Pathak, J.; Hunt, B.; Girvan, M.; Lu, Z.; Ott, E.
\newblock Model-Free Prediction of Large Spatiotemporally Chaotic Systems from
  Data: A Reservoir Computing Approach.
\newblock {\em Phys. Rev. Lett.} {\bf 2018}, {\em 120},~024102.
\newblock
  doi:{\changeurlcolor{black}\href{https://doi.org/10.1103/PhysRevLett.120.024102}{\detokenize{10.1103/PhysRevLett.120.024102}}}.

\bibitem[Schuld and Killoran(2019)]{schuld2019quantum}
Schuld, M.; Killoran, N.
\newblock Quantum machine learning in feature Hilbert spaces.
\newblock {\em Physical Review Letters} {\bf 2019}, {\em 122},~040504.
\newblock
  doi:{\changeurlcolor{black}\href{https://doi.org/10.1103/PhysRevLett.122.040504}{\detokenize{10.1103/PhysRevLett.122.040504}}}.

\bibitem[Abbas \em{et~al.}(2021)Abbas, Sutter, Zoufal, Lucchi, Figalli, and
  Woerner]{Abbas2021}
Abbas, A.; Sutter, D.; Zoufal, C.; Lucchi, A.; Figalli, A.; Woerner, S.
\newblock The power of quantum neural networks.
\newblock {\em Nature Computational Science} {\bf 2021}, {\em 1},~403--409.
\newblock
  doi:{\changeurlcolor{black}\href{https://doi.org/10.1038/s43588-021-00084-1}{\detokenize{10.1038/s43588-021-00084-1}}}.

\bibitem[Nokkala \em{et~al.}(2021)Nokkala, Martínez-Peña, Zambrini, and
  Soriano]{nokkala2021high}
Nokkala, J.; Martínez-Peña, R.; Zambrini, R.; Soriano, M.C.
\newblock High-Performance Reservoir Computing With Fluctuations in Linear
  Networks.
\newblock {\em IEEE Transactions on Neural Networks and Learning Systems} {\bf
  2021}.

\bibitem[Mart{\'\i}nez-Pe{\~n}a \em{et~al.}(2020)Mart{\'\i}nez-Pe{\~n}a,
  Nokkala, Giorgi, Zambrini, and Soriano]{martinez2020information}
Mart{\'\i}nez-Pe{\~n}a, R.; Nokkala, J.; Giorgi, G.L.; Zambrini, R.; Soriano,
  M.C.
\newblock Information Processing Capacity of Spin-Based Quantum Reservoir
  Computing Systems.
\newblock {\em Cognit. Comput.} {\bf 2020}, pp. 1--12.
\newblock
  doi:{\changeurlcolor{black}\href{https://doi.org/10.1007/s12559-020-09772-y}{\detokenize{10.1007/s12559-020-09772-y}}}.

\bibitem[Ghosh \em{et~al.}(2019)Ghosh, Opala, Matuszewski, Paterek, and
  Liew]{Ghosh2019}
Ghosh, S.; Opala, A.; Matuszewski, M.; Paterek, T.; Liew, T.C.H.
\newblock Quantum reservoir processing.
\newblock {\em npj Quantum Inf.} {\bf 2019}, {\em 5},~35.
\newblock
  doi:{\changeurlcolor{black}\href{https://doi.org/10.1038/s41534-019-0149-8}{\detokenize{10.1038/s41534-019-0149-8}}}.

\bibitem[{Ghosh} \em{et~al.}(2020){Ghosh}, {Opala}, {Matuszewski}, {Paterek},
  and {Liew}]{9153954}
{Ghosh}, S.; {Opala}, A.; {Matuszewski}, M.; {Paterek}, T.; {Liew}, T.C.H.
\newblock Reconstructing Quantum States With Quantum Reservoir Networks.
\newblock {\em IEEE Trans. Neural Netw. Learn. Syst.} {\bf 2020}, pp. 1--8.
\newblock
  doi:{\changeurlcolor{black}\href{https://doi.org/10.1109/TNNLS.2020.3009716}{\detokenize{10.1109/TNNLS.2020.3009716}}}.

\bibitem[Ghosh \em{et~al.}(2019)Ghosh, Paterek, and
  Liew]{ghosh2019neuromorphic}
Ghosh, S.; Paterek, T.; Liew, T.C.H.
\newblock Quantum Neuromorphic Platform for Quantum State Preparation.
\newblock {\em Phys. Rev. Lett.} {\bf 2019}, {\em 123},~260404.
\newblock
  doi:{\changeurlcolor{black}\href{https://doi.org/10.1103/PhysRevLett.123.260404}{\detokenize{10.1103/PhysRevLett.123.260404}}}.

\bibitem[Krisnanda \em{et~al.}(2021)Krisnanda, Ghosh, Paterek, and
  Liew]{KRISNANDA2021141}
Krisnanda, T.; Ghosh, S.; Paterek, T.; Liew, T.C.
\newblock Creating and concentrating quantum resource states in noisy
  environments using a quantum neural network.
\newblock {\em Neural Netw.} {\bf 2021}, {\em 136},~141--151.
\newblock
  doi:{\changeurlcolor{black}\href{https://doi.org/https://doi.org/10.1016/j.neunet.2021.01.003}{\detokenize{https://doi.org/10.1016/j.neunet.2021.01.003}}}.

\bibitem[Aspect and Inguscio(2009)]{aspect2009anderson}
Aspect, A.; Inguscio, M.
\newblock Anderson localization of ultracold atoms.
\newblock {\em Phys. Today} {\bf 2009}, {\em 62},~30--35.
\newblock
  doi:{\changeurlcolor{black}\href{https://doi.org/10.1063/1.3206092}{\detokenize{10.1063/1.3206092}}}.

\bibitem[Anderson(1958)]{PhysRev.109.1492}
Anderson, P.W.
\newblock Absence of Diffusion in Certain Random Lattices.
\newblock {\em Phys. Rev.} {\bf 1958}, {\em 109},~1492--1505.
\newblock
  doi:{\changeurlcolor{black}\href{https://doi.org/10.1103/PhysRev.109.1492}{\detokenize{10.1103/PhysRev.109.1492}}}.

\bibitem[Cl{\'{e}}ment \em{et~al.}(2006)Cl{\'{e}}ment, Var{\'{o}}n, Retter,
  Sanchez-Palencia, Aspect, and Bouyer]{Clement_2006}
Cl{\'{e}}ment, D.; Var{\'{o}}n, A.F.; Retter, J.A.; Sanchez-Palencia, L.;
  Aspect, A.; Bouyer, P.
\newblock Experimental study of the transport of coherent interacting
  matter-waves in a 1D random potential induced by laser speckle.
\newblock {\em New Journal of Physics} {\bf 2006}, {\em 8},~165--165.
\newblock
  doi:{\changeurlcolor{black}\href{https://doi.org/10.1088/1367-2630/8/8/165}{\detokenize{10.1088/1367-2630/8/8/165}}}.

\bibitem[Billy \em{et~al.}(2008)Billy, Josse, Zuo, Bernard, Hambrecht, Lugan,
  Cl{\'e}ment, Sanchez-Palencia, Bouyer, and Aspect]{billy2008direct}
Billy, J.; Josse, V.; Zuo, Z.; Bernard, A.; Hambrecht, B.; Lugan, P.;
  Cl{\'e}ment, D.; Sanchez-Palencia, L.; Bouyer, P.; Aspect, A.
\newblock {Direct observation of Anderson localization of matter waves in a
  controlled disorder}.
\newblock {\em Nature} {\bf 2008}, {\em 453},~891--894.
\newblock
  doi:{\changeurlcolor{black}\href{https://doi.org/10.1038/nature07000}{\detokenize{10.1038/nature07000}}}.

\bibitem[Roati \em{et~al.}(2008)Roati, D'Errico, Fallani, Fattori, Fort,
  Zaccanti, Modugno, Modugno, and Inguscio]{roati2008anderson}
Roati, G.; D'Errico, C.; Fallani, L.; Fattori, M.; Fort, C.; Zaccanti, M.;
  Modugno, G.; Modugno, M.; Inguscio, M.
\newblock {Anderson localization of a non-interacting Bose--Einstein
  condensate}.
\newblock {\em Nature} {\bf 2008}, {\em 453},~895--898.
\newblock
  doi:{\changeurlcolor{black}\href{https://doi.org/10.1038/nature07071}{\detokenize{10.1038/nature07071}}}.

\bibitem[Modugno(2006)]{PhysRevA.73.013606}
Modugno, M.
\newblock Collective dynamics and expansion of a Bose-Einstein condensate in a
  random potential.
\newblock {\em Phys. Rev. A} {\bf 2006}, {\em 73},~013606.
\newblock
  doi:{\changeurlcolor{black}\href{https://doi.org/10.1103/PhysRevA.73.013606}{\detokenize{10.1103/PhysRevA.73.013606}}}.

\bibitem[Huntley(1989)]{ley1989specklhunte}
Huntley, J.
\newblock Speckle photography fringe analysis: assessment of current
  algorithms.
\newblock {\em Appl. Opt.} {\bf 1989}, {\em 28},~4316--4322.
\newblock
  doi:{\changeurlcolor{black}\href{https://doi.org/10.1364/AO.28.004316}{\detokenize{10.1364/AO.28.004316}}}.

\bibitem[Mujal \em{et~al.}(2019)Mujal, Polls, Pilati, and
  Juli\'a-D\'{\i}az]{PhysRevA.100.013603}
Mujal, P.; Polls, A.; Pilati, S.; Juli\'a-D\'{\i}az, B.
\newblock Few-boson localization in a continuum with speckle disorder.
\newblock {\em Phys. Rev. A} {\bf 2019}, {\em 100},~013603.
\newblock
  doi:{\changeurlcolor{black}\href{https://doi.org/10.1103/PhysRevA.100.013603}{\detokenize{10.1103/PhysRevA.100.013603}}}.

\bibitem[Mujal \em{et~al.}(2020)Mujal, Mart\'{i}nez~Miguel, Polls,
  Juliá-Díaz, and Pilati]{zenodo}
Mujal, P.; Mart\'{i}nez~Miguel, A.; Polls, A.; Juliá-Díaz, B.; Pilati, S.
\newblock Database used for the supervised learning of few dirty bosons with
  variable particle number.
\newblock {\em Zenodo} {\bf 2020}.
\newblock
  doi:{\changeurlcolor{black}\href{https://doi.org/10.5281/zenodo.4058492}{\detokenize{10.5281/zenodo.4058492}}}.

\bibitem[Mujal(2019)]{pmtphdthesis}
Mujal, P.
\newblock
  \href{http://brunojulia.fqa.ub.edu/works/PMT_phD_Thesis_book.pdf}{Interacting
  ultracold few-boson systems}.
\newblock PhD thesis, Universitat de Barcelona,  2019.

\bibitem[Mart\'{\i}nez-Pe\~na \em{et~al.}(2021)Mart\'{\i}nez-Pe\~na, Giorgi,
  Nokkala, Soriano, and Zambrini]{martinez2021dynamical}
Mart\'{\i}nez-Pe\~na, R.; Giorgi, G.L.; Nokkala, J.; Soriano, M.C.; Zambrini,
  R.
\newblock Dynamical Phase Transitions in Quantum Reservoir Computing.
\newblock {\em Phys. Rev. Lett.} {\bf 2021}, {\em 127},~100502.
\newblock
  doi:{\changeurlcolor{black}\href{https://doi.org/10.1103/PhysRevLett.127.100502}{\detokenize{10.1103/PhysRevLett.127.100502}}}.

\bibitem[Kutvonen \em{et~al.}(2020)Kutvonen, Fujii, and Sagawa]{Kutvonen2020}
Kutvonen, A.; Fujii, K.; Sagawa, T.
\newblock Optimizing a quantum reservoir computer for time series prediction.
\newblock {\em Sci. Rep.} {\bf 2020}, {\em 10},~14687.
\newblock
  doi:{\changeurlcolor{black}\href{https://doi.org/10.1038/s41598-020-71673-9}{\detokenize{10.1038/s41598-020-71673-9}}}.

\bibitem[Chen and Nurdin(2019)]{Chen2019}
Chen, J.; Nurdin, H.I.
\newblock Learning nonlinear input--output maps with dissipative quantum
  systems.
\newblock {\em Quantum Inf. Process.} {\bf 2019}, {\em 18},~198.
\newblock
  doi:{\changeurlcolor{black}\href{https://doi.org/10.1007/s11128-019-2311-9}{\detokenize{10.1007/s11128-019-2311-9}}}.

\bibitem[Nakajima \em{et~al.}(2019)Nakajima, Fujii, Negoro, Mitarai, and
  Kitagawa]{PhysRevApplied.11.034021}
Nakajima, K.; Fujii, K.; Negoro, M.; Mitarai, K.; Kitagawa, M.
\newblock Boosting Computational Power through Spatial Multiplexing in Quantum
  Reservoir Computing.
\newblock {\em Phys. Rev. Applied} {\bf 2019}, {\em 11},~034021.
\newblock
  doi:{\changeurlcolor{black}\href{https://doi.org/10.1103/PhysRevApplied.11.034021}{\detokenize{10.1103/PhysRevApplied.11.034021}}}.

\bibitem[Mujal \em{et~al.}(2021)Mujal, Nokkala, Mart{\'{\i}}nez-Pe{\~{n}}a,
  Giorgi, Soriano, and Zambrini]{Mujal_2021nonlinearities}
Mujal, P.; Nokkala, J.; Mart{\'{\i}}nez-Pe{\~{n}}a, R.; Giorgi, G.L.; Soriano,
  M.C.; Zambrini, R.
\newblock Analytical evidence of nonlinearity in qubits and continuous-variable
  quantum reservoir computing.
\newblock {\em Journal of Physics: Complexity} {\bf 2021}, {\em 2},~045008.
\newblock
  doi:{\changeurlcolor{black}\href{https://doi.org/10.1088/2632-072x/ac340e}{\detokenize{10.1088/2632-072x/ac340e}}}.

\bibitem[Tran and Nakajima(2020)]{higherorderqrc}
Tran, Q.H.; Nakajima, K.
\newblock Higher-Order Quantum Reservoir Computing.
\newblock (Preprint) arXiv:2006.08999.

\end{thebibliography}
\end{document}